# Signal detection algorithms for interferometric sensors with harmonic phase modulation: distortion analysis and suppression

Leonid Liokumovich[1], Andrei Medvedev[1], Konstantin Muravyov[1], Philipp Skliarov[1], Nikolai Ushakov[1,2,*]

[1]*Department of Radiophysics, Peter the Great St. Petersburg Polytechnic University, 29, Polytechnicheskaya ul., St. Petersburg, 195251, Russia*
[2]*AstroSoft, Gelsingforsskaya 3, corpus 11D, St. Petersburg, 194044, Russia*
*\*Corresponding author: n.ushakoff@spbstu.ru*

**In current paper, distortions in digital demodulation schemes with harmonic phase modulation for interferometric optical sensors are considered. In particular, the influence of target signal variations on phase demodulation errors is theoretically evaluated. An analytical expression, describing the phase error magnitude dependence on first derivative and mean value of measured signal and amplitude of the phase modulation in case of simple 4-point demodulation algorithm is derived. After that, an approach for synthesizing algorithms with suppressed sensitivity to target signal variations is developed. Based on this approach, a novel 4+1 demodulation algorithm is proposed. It is shown analytically that the demodulation error of new 4+1 algorithm is proportional to the second derivative of target signal, and therefore, is typically several orders of magnitude smaller than in case of 4-point algorithm. The correspondence between analytical expressions and real phase errors, induced by target signal variations is verified by means of numeric simulation.**

http://dx.doi.org/10.1364/AO.99.099999

## 1. INTRODUCTION

Optical fiber sensors gain extensive attention from both academia and industry during past several decades. Among others, interferometric sensors have ability to perform measurement of various physical quantities, offering high resolution and dynamic range, ability for remote sensing and to perform in harsh environments.

Interferometric measurements inevitably require demodulation approach, extracting the target phase difference from interference signal $u(t)$. One of the most widely used and efficient methods utilizes auxiliary phase modulation of interference signal argument. This modulation is used in various interferometric systems, sometimes optical schemes are intentionally modified in order to allow introduction of phase modulation [1–7]. In case of digital representation of interference signal with sampling frequency $f_d$, one has to deal with stream of samples of the following form

$$u_i = u(t_i) = U_0 + U_m \cos(\varphi + \psi_i), \qquad (1)$$

where $\psi_i = \psi(t_i)$, $t_i = i/f_d$, $\varphi$ – target phase difference, carrying information about the actual measurand, $U_0$ and $U_m$ – constant part and amplitude of interference signal, $\psi(t)$ – known periodical signal of phase modulation with frequency $f_M$, $i$ – sample number. As follows from technical feasibility, modulation and sampling frequencies are chosen such that $f_d = N f_M$ is integer multiple of $f_M$ and $N$ is number of samples on modulation period. Generally, parameters $U_0$, $U_m$ can vary with respect to time due to fluctuations in optical setup as well as $\varphi(t)$ is non-stationary due to changes of target signal. However, spectrum width of $\varphi(t)$ signal is much lower than the frequencies $f_d$ and $f_M$.

Initially, interferometric systems utilized analog modulation and demodulation methods [1,8-10], although, some of them are used nowadays. However, most state of the art interferometric systems utilize digital demodulation techniques, allowing more flexibility to algorithm design [2,3,7,11-14]. Such techniques generally imply some demodulation interval, formed by $Q$ signal samples. In most cases it is assumed that parameters $U_0$, $U_m$ and $\varphi$ are constant on this interval and can be found (or just the latter one) as a solution of a system of linear equations (SLE)

$$\begin{cases} u^{(0)} = U_0 + U_m \cdot \cos(\varphi + \psi^{(0)}); \\ u^{(1)} = U_0 + U_m \cdot \cos(\varphi + \psi^{(1)}); \\ \dots \\ u^{(Q-1)} = U_0 + U_m \cdot \cos(\varphi + \psi^{(Q-1)}), \end{cases} \quad (2)$$

where $u^{(q)}$, $q = 0, 1, 2 \dots Q-1$ are samples of interference signal. In case of $Q=3$ the system has a unique exact solution for three unknowns $U_0$, $U_m$ and $\varphi$. Otherwise, in case of $Q > 3$, an approximate solution of eq. (2) must be found, for example, by means of least squares approach (see Appendix 1). The ensemble of $\psi^{(q)}$, determined by $\psi(t_q)$ signal and equation, used for solving eq. (2) is a basis of demodulation approach. Moreover, demodulation approach implies positions of demodulation intervals with respect to the whole signal $u(t)$ and periods of phase modulation. In simplest and most feasible case demodulation intervals are offset from one another at $kN$ samples ($k$ – integer). Otherwise different equations for calculating $\varphi$ from $u^{(q)}$, corresponding to different sets $\psi^{(q)}$ will be needed.

A vast number of references, dealing with such approaches, are dedicated to interferogram analysis and processing [1–3,7,11,15,16], where the target phase is a two-dimensional function of two coordinates, however, it is stationary with respect to time. In this case, samples of signal (1) become two-dimensional interference pictures; however, the initial calculation of phase in every point is performed in the same manner. In most cases $\psi(t_q)$ is linear function of time with fixed increment $\pi/2$ between adjacent samples, resulting in most simple and convenient form of solutions of eq. (2). However, application of these principles to interferometric sensors, when the measurand varies with respect to time [17] is somehow different.

First of all, in sensors, registering temporal signals, reproducibility of $\psi^{(q)}$ sets is needed, requiring periodic character of signal $\psi(t)$. A direct analogy to linearly increasing $\psi(t)$ is saw-tooth periodic modulation with amplitude $2\pi$. However, due to various reasons other types of modulation, such as harmonic may be attractive and even preferable. Despite the fact that phase harmonic modulation is widely used in interferometric measurements, the theory and analysis are not enough developed. In most cases consideration is limited to specialized cases, for example, in [18] an algorithm with $N=12$ and fixed amplitude is developed, in [19,20] an algorithm with minimized influence of additive noises is developed. However, in known literature equations for calculating phase in closed form in general case of arbitrary $N$ are absent.

Secondly, target phase signal $\varphi(t)$ is variable with respect to time by its nature. Therefore, calculating phase according to solution of eq. (2) will yield an erroneous result, distorted with respect to the true phase, which requires additional analysis and measures in order to decrease the obtained distortion.

In the current paper, an improved algorithm, enabling one to decrease the distortion, produced by variation of target phase on demodulation interval, is presented.

## 2. Simple algorithm with harmonic modulation

In general, harmonic modulation signal can be written in a form

$$\psi_i = \psi(t_i) = \psi_m \sin(2\pi f_M t_i + \theta) = \psi_m \sin\left[\frac{2\pi}{N} \cdot i + \theta\right], \quad (3)$$

where $\psi_m$ and $\theta$ are amplitude and initial phase of modulation signal. In case of arbitrary $\theta$ analysis becomes quite complicated, therefore, for certainty we'll consider the simplest for practical implementation case $\theta=0$.

In case when the first points of demodulation interval and modulation period coincide, modulation samples can be written as

$$\psi^{(q)} = \psi_m \sin\left(\frac{2\pi}{N} \cdot q\right), q = 0, 1, \dots, Q-1. \quad (4)$$

As was already mentioned, interference signal (1) has at least three parameters: $\varphi$, $U_0$ and $U_m$, which are assumed to be unknown, and therefore, at least three equations, forming a nondegenerate SLE in eq. (2) are required. The nondegeneracy condition in this case is inequality of at least three $\psi^{(q)}$ values by modulo $2\pi$. Therefore, the shortest possible demodulation interval corresponds to $Q=N=3$ and each single value of target phase $\varphi$ is calculated based on $u^{(0)}$, $u^{(1)}$ and $u^{(2)}$. However, the ratio $f_d/f_M=3$ is sometimes technically inconvenient and therefore, the case $N=4$ occurs to be more practical. Then in case of demodulation interval coinciding with modulation period, $Q=N=4$, $\psi^{(0)} = \psi^{(2)} = 0$, $\psi^{(1)} = -\psi^{(3)} = \psi_m$, and according to solution of SLE (2) in Appendix 1 by means of ordinary least squares (OLS) approach, coefficients in eq. (A1-4) can be written as

$$\begin{aligned} A_{11} = A_{12} = A_{23} = 0, \quad A_{13} = 4 \cdot [1 - \cos(\psi_m)]^2, \\ A_{21} = -4 \cdot [1 + \cos(\psi_m)] \sin^2(\psi_m), A_{22} = 8 \cdot \sin^2(\psi_m). \end{aligned} \quad (5)$$

Substituting coefficients from eq. (5) to eq. (A1-3) and making several simplifications, one gets the following simple equation for target phase calculation

$$\varphi_r = -\mathrm{atan2}\left[\frac{1-\cos(\psi_m)}{\sin(\psi_m)} \cdot \frac{u^{(1)} - u^{(3)}}{u^{(0)} - u^{(1)} + u^{(2)} - u^{(3)}}\right]. \quad (6)$$

The function atan2 is more commonly written as a function of two arguments atan2($y$, $x$), where $y$ and $x$ are nominator and denominator of atan function. The atan2 function takes into account signs of $x$ and $y$ and therefore returns $\varphi_r$ within [–π, π) range. In this and consequent equations for target phase calculation the factor [1– cos($\psi_m$)]/sin($\psi_m$) must be treated as a part of nominator. It also should be noted that $\psi_m$ has limited range space in order to avoid nondegeneracy of SLE in eq. (2). The demodulation approach, based on eq. (6), will further be denoted as OLS-4. Although eq. (6) is a special case of a well-known general OLS solution of system (2) in case of harmonic modulation (4) and $N=Q=4$, it isn't presented in explicit form in any known literature.

In further sections we will consider the problem of demodulated phase distortion caused by signal phase variations on demodulation interval (also referred to as dynamic distortion). This problem is relevant for most of the algorithms for interference signal demodulation, however, in order to be specific, we will consider the case of $N=4$ and algorithm (6).

## 3. Signal distortion, induced by temporal variations of target phase

Solutions of eq. (2) in eqs. (A1-3) and (6) were obtained under assumption of stationary target phase $\varphi$. However, in practical interferometric sensors $\varphi$ contains information about the actual measurand and therefore, inevitably changes with respect to time. Strictly, each point of interference signal corresponds to particular value of target phase $\varphi^{(q)}$, varying with respect to $q$. Therefore, generally, calculated value $\varphi_r$ will not be equal to any of the $\varphi^{(q)}$. In order to learn how the target phase change on the demodulation interval will affect $\varphi_r$, one needs to find a relation between phase variation $\varphi^{(q)}$ and $\varphi_0$ – a value to be considered as true target phase on demodulation interval and then analyze the phase error $\Delta\varphi = \varphi_r - \varphi_0$ for every considered demodulation algorithm.

If the frequency band of target phase oscillations is much smaller than $f_M$ and target phase increment on demodulation interval (with length $1/f_M$) is small, approximate analytical consideration is possible. For that one needs to linearize target phase change in close vicinity of some point $t_0$ in form $\varphi(t) \approx \varphi_0 + [d\varphi(t)/dt]\cdot(t-t_0)$, where $\varphi_0 = \varphi(t_0)$ and derivative value at point $t = t_0$ is taken. Since $Q$ samples of interference signal are used for target phase calculation, it is convenient to relate the point $t = t_0$ with the middle of the interval between moments at which $u^{(0)}$ and $u^{(Q-1)}$ samples were taken. In system of discrete samples linear phase change on demodulation interval can be characterized by increment between adjacent samples $\delta = [\varphi^{(Q-1)} - \varphi^{(0)}]/Q = [d\varphi(t)/dt]/f_d$, then the target phase samples can be written as

$$\varphi^{(q)} = \varphi_0 + \delta \cdot [q - (Q-1)/2], \qquad (7)$$

where $\varphi_0$ can be expressed in terms of $\varphi^{(q)}$ samples as

$$\varphi_0 = \varphi^{\left(\frac{Q-1}{2}\right)}, \textbf{ odd } Q; \quad \varphi_0 = 0.5\cdot\left(\varphi^{\left(\frac{Q}{2}\right)} + \varphi^{\left(\frac{Q}{2}-1\right)}\right), \textbf{ even } Q. \qquad (8)$$

Therefore, instead of considering fixed target phase value on demodulation interval, we consider an approximate of true phase value $\varphi_0$ and small parameter $\delta \ll 1$, characterizing target phase derivative.

Using representation of phase samples in form of (7) and (8), deviations of real samples of signal $u(t)$ from ideal form (2) can be considered and the resultant difference between demodulated value $\varphi_r$ and true value $\varphi_0$ can be evaluated analytically. Let us demonstrate such result in case of $N=4$ and algorithm (6). In this case according to eqs. (7) and (8), samples of interference signals can be written as

$$u^{(0)} = U_0 + U_m \cos\left(\varphi_0 - \frac{3}{2}\delta\right), \quad u^{(1)} = U_0 + U_m \cos\left(\varphi_0 - \frac{1}{2}\delta + \psi_m\right),$$
$$u^{(2)} = U_0 + U_m \cos\left(\varphi_0 + \frac{1}{2}\delta\right), \quad u^{(3)} = U_0 + U_m \cos\left(\varphi_0 + \frac{3}{2}\delta - \psi_m\right). \qquad (9)$$

In case of target phase being constant during the demodulation interval, the eq. (6) is correct and $\varphi_r = \varphi_0 = \operatorname{atan2}(S/C)$, where $S$ and $C$ are proportional to sin and cos of $\varphi_0$ with the same factor of proportionality. Coefficients $C$ and $S$ are found from samples of interference signal according to eq. (6) under assumptions $\varphi^{(q)} = \varphi_0$ and $\delta = 0$. However, if one plugs $u^{(q)}$ samples, accounting for changing $\varphi^{(q)}$ in case of nonzero $\delta$ into eq. (6), coefficients $S_r$ and $C_r$ are obtained. It should be noted that $S_r$ and $C_r$ are dependent on $\delta$ and generally different from $S$ and $C$. Therefore, calculated phase value $\varphi_r = \operatorname{atan2}(S_r/C_r)$ also depends on $\delta$ and generally is different from $\varphi_0$. Under assumption of small phase error, one can expand $\varphi_r$ into Tailor series with respect to $\delta$. Neglecting the terms with order of smallness greater than first, the phase error $\Delta\varphi$ can be written as

$$\Delta\varphi \approx \frac{d}{d\delta}\operatorname{atan2}\left(\frac{S_r}{C_r}\right)\bigg|_{\delta=0} \cdot \delta = \delta\frac{S'C - C'S}{C^2 + S^2} =$$
$$\delta\frac{S'\cos(\varphi_0) - C'\sin(\varphi_0)}{2(1 - \cos(\psi_m))}, \qquad (10)$$

where expressions for $S_r$, $C_r$, $S$ and $C$ as well as for derivatives of $S_r$ and $C_r$ with respect to $\delta$, denoted as $S'$ and $C'$ are given in Appendix 3. Substituting eqs. (A3-1) – (A3-3) into eq. (10) and making several simplifications, the following equation for phase error is obtained

$$\Delta\varphi \approx \frac{\delta}{2}\left[\frac{\cos(2\varphi_0 - \psi_m/2)}{\sin(\psi_m/2)\cdot\sin(\psi_m)} - \frac{\cos(\psi_m)}{1 - \cos(\psi_m)}\right]. \qquad (11)$$

Eq. (11) enables one to estimate the value of phase demodulation error for a target signal with given parameters $\varphi_0$ and $\delta$, phase modulation with amplitude $\psi_m$ and in case when eq. (6) is used for phase calculation. Detailed analysis of eq. (11) will be presented below, however, even from a brief look it can be seen that the error $\Delta\varphi$ increases as modulation amplitude $\psi_m$ approaches zero or a multiple of $\pi$. Also, for each $\psi_m$ value, values of $\varphi_0$ exist, minimizing and maximizing the resultant error.

## 4. Synthesis of algorithms with reduced phase errors

Since only three samples of interference signal are required for unambiguous calculation of target phase $\varphi_r$ in case of different $\psi^{(q)}$ values, the rest samples are redundant and may be used to extract additional information about the phase signal. Therefore, algorithms with $Q>3$ can be used in order to reduce the resulting phase error $\Delta\varphi$. Error reduction will take place if the interference signal's samples $u^{(q)}$ correspond to different phase shifts (it should be noted that target phase shifts also are taken into account, therefore, terms $\psi^{(q)} + \delta \cdot [q - (Q-1)/2]$ must be different for different $q$). As before, inequality is considered in terms of modulo $2\pi$.

Two approaches for reduction of error, caused by quasilinear variation of target phase, can be proposed. The first one is based on solving the system of trigonometric equations with four variables $\varphi_0$, $U_0$, $U_m$ and $\delta$. This approach enables one to provide full rejection of the phase error, however, generally it is quite complicated and may even not have an analytical solution. Examples of algorithms developed with the use of such approach in case of linear phase modulation and for suppression of other distortions, can be found in [3,7,21].

In case of harmonic phase modulation, another approach is more efficient. It doesn't provide total rejection of the phase error, however, enables one to sufficiently reduce it, in particular – the phase error becomes proportional to high order powers of $\delta$ and high order derivatives of target signal. This approach is much more successful than the first one and will be utilized below.

This approach was implemented in different ways in [3,12,21]. The most simple and general way was proposed in [22]. Although [22] was dedicated to linear modulation and analysis of different type of distortions, this approach can be applied in other cases, including our task.

Let us assume that the equation for calculating target phase has the following form

$$\tan(\varphi_r) = \frac{a_0 u^{(0)} + a_1 u^{(1)} + \ldots + a_{Q-1} u^{(Q-1)}}{b_0 u^{(0)} + b_1 u^{(1)} + \ldots + b_{Q-1} u^{(Q-1)}}, \qquad (12)$$

where $a_q$ and $b_q$ – fixed coefficients. Substituting expressions for some particular $u^{(q)}$ for given $Q$ and phase modulation format, one can obtain expressions, proportional to $\sin(\varphi)$ and $\cos(\varphi)$ in nominator and denominator, respectively. For correct performance of atan2 function, proportionality factors must be equal and both positive. According to these requirements, $a_q$ and $b_q$ coefficients must fulfill the following conditions

$$\sum_{q=0}^{Q-1} a_q = 0, \quad \sum_{q=0}^{Q-1} a_q \cos(\psi^{(q)}) = 0, \quad \sum_{q=0}^{Q-1} b_q = 0, \quad \sum_{q=0}^{Q-1} b_q \sin(\psi^{(q)}) = 0,$$
$$-\sum_{q=0}^{Q-1} a_q \sin(\psi^{(q)}) = \sum_{q=0}^{Q-1} b_q \cos(\psi^{(q)}) > 0. \qquad (13)$$

It can be seen from eq. (12) that it will still hold for all $a_q$, $b_q$ multiplied at the same nonzero number. Therefore, the value of any $a_q$ or $b_q$ coefficient can be set arbitrarily and the rest coefficients – expressed in terms of it. In such a manner, for any particular $Q$, one must find $2Q-1$ unknown coefficients, while eq. (13) provides only five equations. Therefore, some voluntarism takes place in process of $a_q$, $b_q$ coefficients selection, enabling a possibility of algorithm synthesis under additional conditions on the algorithm properties.

In order to reduce the phase error $\Delta\varphi$, additional conditions on $a_q$ and $b_q$ coefficients, such that different particular components of error become equal to zero. For that, an expression, analogous to eq. (11) must be derived from eq. (12) and set to zero, defining additional conditions on $a_q$ and $b_q$ coefficients. Coefficients for the sought error-suppressed algorithm can be found from the resulting equations.

This generalized approach can be used for synthesis of particular algorithms, reducing specified distortion mechanisms and working in case of given phase modulation mode. As such particular case, let us consider dynamic distortion, caused by linear increment of target phase and harmonic phase modulation given by eq. (4), $N=4$.

It can be seen in eq. (9) that in case of $N=Q=4$ we have four samples with different phase shifts, indicating an ability to obtain error-reduction algorithm. However, as shown in Appendix 2, the resulting structure of signal samples doesn't fulfill all arising conditions and therefore, error-reduction algorithm for these parameters can't be synthesized. In spite of this, let us consider ability of error-compensation in case of $N=4$ and $Q=5$, i.e., when four samples from the current modulation period and one sample from next modulation period are processed in order to find the target phase. Algorithms, using such set of samples are often denoted as $N+1$ [22–24]. The next section is devoted to design and analysis of 4+1 algorithm, which can be considered as an enhanced alternative of algorithm (6).

## 5. 4+1 algorithm with reduced phase errors

In case of phase modulation given be eq. (4) with $N=4$ and $Q=5$ and taking into account quasi-linear variation of target phase according to eqs. (7) and (8), registered samples of interference signal are written as follows

$$u^{(0)} = U_0 + U_m \cos(\varphi_0 - 2\delta), \; u^{(1)} = U_0 + U_m \cos(\varphi_0 - \delta + \psi_m),$$
$$u^{(2)} = U_0 + U_m \cos(\varphi_0), \; u^{(3)} = U_0 + U_m \cos(\varphi_0 + \delta - \psi_m),$$
$$u^{(4)} = U_0 + U_m \cos(\varphi_0 + 2\delta).$$
(14)

Since demodulation interval includes 5 samples, center $t_0$ corresponds to sample $u^{(2)}$ and $\varphi_0 = \varphi^{(2)}$. Five signal samples in eq. (14) have different phase shifts and hence, provide an ability to compensate two error components, for example, the ones proportional to $\delta$ and $\delta^2$. For harmonic modulation given by eq. (4) in case of $N=4$ and $Q=5$ eq. (13) will be rewritten as

$$a_0+a_1+a_2+a_3+a_4=0, \; a_0+a_2+a_4+(a_1+a_3)\cdot\cos(\psi_m)=0,$$
$$b_0+b_1+b_2+b_3+b_4=0, \; (-b_1+b_3)\cdot\sin(\psi_m)=0,$$
$$(-a_1+a_3)\cdot\sin(\psi_m)=b_0+b_2+b_4+(b_1+b_3)\cdot\cos(\psi_m)>0.$$
(15)

Further we find an expression for phase error $\Delta\varphi$ in case of demodulation performed according to eq. (12) from interference signal samples set given by eq. (14). This can be done in a similar manner to the way eq. (11) was derived, substituting samples in eq. (14) into eq. (12), expanding $\varphi_r = \text{atan2}(S_r/C_r)$ into Tailor series with respect to $\delta$ and omitting all terms smaller than third order of magnitude, resulting in the following approximate equation

$$\Delta\varphi \approx \frac{d}{d\delta}\text{atan2}\left(\frac{S_r}{C_r}\right)\bigg|_{\delta=0}\cdot\delta + \frac{1}{2}\frac{d^2}{d\delta^2}\text{atan2}\left(\frac{S_r}{C_r}\right)\bigg|_{\delta=0}\cdot\delta^2,$$
(16)

Using first and second derivatives of $S_r$ and $C_r$ with respect to $\delta$, denoted as $S'$, $C'$, $S''$ and $C''$, introduced in Appendix 3, eq. (16) can be rewritten as

$$\Delta\varphi \approx \delta\frac{S'C-C'S}{C^2+S^2} + \frac{1}{2}\delta^2\frac{(S''C-C''S)(C^2+S^2)-2(CC'+SS')(S'C-C'S)}{(C^2+S^2)^2}.$$
(17)

In order to ensure the condition $\Delta\varphi=0$, one needs to find such coefficients $a_q$ and $b_q$, that both terms in eq. (17) are equal to zero. Substituting eqs. (A3-5), (A3-6) and (A3-8) into eq. (17), one obtains the following condition

$$\frac{(A_1+A_2)\sin(2\varphi_0)+(B_1+B_2)\cos(2\varphi_0)+(D_1+D_2)}{b_0+b_2+(b_1+b_3)\cos(\psi_m)}=0,$$
(18)

where

$$A_1 = \delta\cdot[2(a_0-a_4)+(a_1-a_3)\cos(\psi_m)-(b_1+b_3)\sin(\psi_m)],$$
$$B_1 = \delta\cdot[(a_1+a_3)\sin(\psi_m)+2(b_0-b_4)+(b_1-b_3)\cos(\psi_m)],$$
$$D_1 = \delta\cdot[(a_1+a_3)\sin(\psi_m)-2(b_0-b_4)-(b_1-b_3)\cos(\psi_m)],$$
$$A_2 = \delta^2\cdot[0.5(a_1-a_3)\sin(\psi_m)+2(b_0+b_4)+0.5(b_1+b_3)\cos(\psi_m)],$$
$$B_2 = \delta^2\cdot[-2(a_0+a_4)-0.5(a_1+a_3)\cos(\psi_m)+0.5(b_1-b_3)\sin(\psi_m)],$$
$$D_2 = \delta^2\cdot[-2(a_0+a_4)-0.5(a_1+a_3)\cos(\psi_m)-0.5(b_1-b_3)\sin(\psi_m)].$$
(19)

Fulfillment of conditions $A_1=0$, $B_1=0$, $D_1=0$ and $A_2=0$, $B_2=0$, $D_2=0$ eliminates the error components, proportional to $\delta$ and $\delta^2$, respectively. As a result of eqs. (15) and (19), we get eleven equations and nine variables. However, it can be shown that all eleven equations have a solution in case

$$a_0 = -a_4 = \frac{1}{2}a_1, \; a_2 = 0, \; a_3 = -a_1,$$
$$b_1 = b_3 = a_1\cdot\frac{\sin(\psi_m)}{1-\cos(\psi_m)}, \; b_0 = b_4 = -\frac{1}{4}b_1, \; b_2 = -\frac{3}{2}b_1.$$
(20)

It was taken into account that one of the coefficients ($a_1$ in this case) can be chosen arbitrarily. Taking into account the condition on the signs of nominator and denominator in eq. (12), the choice $a_1 = -4[1-\cos(\psi_m)]/\sin(\psi_m)$ occurs to be feasible, leading to the following equation for target phase calculation

$$\varphi_r = -\text{atan2}\left[\frac{1-\cos(\psi_m)}{\sin(\psi_m)}\cdot\frac{2u^{(0)}+4u^{(1)}-4u^{(3)}-2u^{(4)}}{u^{(0)}-4u^{(1)}+6u^{(2)}-4u^{(3)}+u^{(4)}}\right].$$
(21)

The comments to eq. (6) below it hold for eq. (21) as well.

As a result of suppressing error components, proportional to $\delta$ and $\delta^2$, residual phase error $\Delta\varphi$ for the developed 4+1 algorithm will be primarily determined by second derivative of target phase signal, particularly, by parameter $\gamma = [d^2\varphi(t)/dt^2]/f_d^2$ (derivative value at time moment $t_0$ is implied, $t_0$ being the center of interval between $u^{(0)}$ and $u^{(Q-1)}$). Taking into account second derivative, target signal can be written as

$$\varphi^{(q)} = \varphi_0 + \delta\cdot[q-(Q-1)/2] + \gamma\cdot[q-(Q-1)/2]^2/2.$$
(22)

Plugging eq. (22) into eq. (2) in case of $Q=5$, one obtains

$$u^{(0)} = U_0 + U_m \cos(\varphi_0 - 2\delta + 2\gamma),$$
$$u^{(1)} = U_0 + U_m \cos\left(\varphi_0 - \delta + \frac{1}{2}\gamma + \psi_m\right),$$
$$u^{(2)} = U_0 + U_m \cos(\varphi_0), \qquad (23)$$
$$u^{(3)} = U_0 + U_m \cos\left(\varphi_0 + \delta + \frac{1}{2}\gamma - \psi_m\right),$$
$$u^{(4)} = U_0 + U_m \cos(\varphi_0 + 2\delta + 2\gamma).$$

In the same way as eqs. (14) and (19) were derived, one can estimate the target phase error in case of algorithm based on eq. (21). Considering the error component proportional to $\gamma$, one obtains the following

$$\Delta\varphi \approx \frac{\partial}{\partial\gamma} \mathrm{atan}2\left(\frac{S_r}{C_r}\right) = \frac{1}{2}\gamma. \qquad (24)$$

By this means, the proposed 4+1 algorithm produces a residual phase error, which is proportional to $\gamma$, but is independent of both $\varphi_0$ and $\psi_m$. In case of $\psi_m$ close to multiple of $\pi$, on the one hand, phase error $\Delta\varphi$ increases, which will be shown in analysis below and on the other hand, eq. (24) becomes inadequate in terms of describing the phase error.

## 6. Numeric simulation

In order to test the algorithms (6) and (21) and verify the phase error estimates given by eqs. (11) and (24), a numeric modeling was carried out. Modeling was chosen to examine the proposed algorithms instead of live experiment due to several reasons:

- in most current interference systems the observed signal is much alike the one in eq. (1)
- for those noises and distortions, present in the experimental interference signal, it is not always easy to determine where they originate from.

Therefore, as the aim of this paper is to investigate the influence of target phase variation during demodulation interval on the resultant phase error, simulation is a better choice for testing the developed theory since it doesn't produce any other error sources.

The modeling was performed in the following way – interference signal was modeled according to eq. (1), where the amplitude of phase modulation $\psi_m$ was varied within the interval $(0, \pi)$. In the presented examples, a practical value of modulation frequency 25 kHz was used, corresponding to sampling frequency 100 kHz. The signals were calculated for a time interval 0.1 seconds length. Target phase signal $\varphi$ was chosen to be harmonic signal in order to simplify the evaluation of phase error magnitude. Frequency and amplitude of target signal $\varphi$ were varied within [10 - 1000] Hz and [0.1 - 10] rad, respectively. An example of interference signal on time interval 0.2 - 0.6 ms is shown in figure 1.

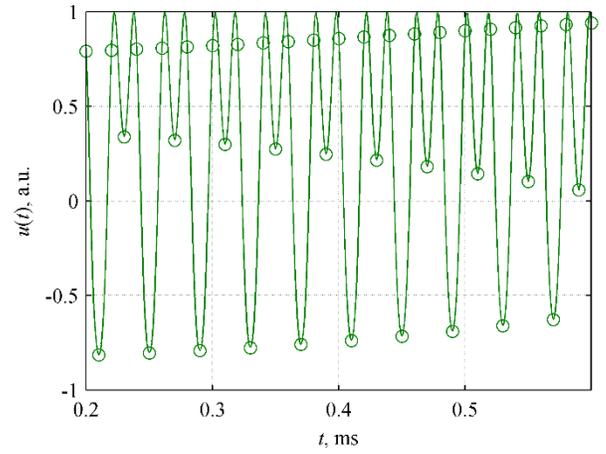

Figure 1. Fragment of interference signal in case of sampling frequency 100 kHz, modulation frequency 25 kHz, modulation amplitude 1.8 rad.

Points correspond to signal samples $u_i$, while the solid curve illustrates continuous signal shape.

Such signals were calculated for all variants of modulation amplitude and signal properties (frequency, amplitude and constant component). After that, two above-mentioned algorithms described by eqs. (6) and (21) were applied to simulated interference signals, producing two demodulated signals. Demodulation errors were estimated at each demodulation interval (for each sample of demodulated signal) as differences between the values of initial target signal at center of demodulation interval and demodulated signal samples. In figure 2 comparison of demodulated signal $\varphi_r$ and errors in case of two algorithms $\Delta\varphi_{OLS-4}$ and $\Delta\varphi_{4+1}$ is shown for target signal with 5 radians amplitude and 100 Hz frequency. Modulation amplitude $\psi_m$ was 0.7 radians and constant component of target phase $\varphi_0 = 4$ radians (subtracted for $\varphi_r$).

For the same target signal, a comparison of phase errors as functions of modulation amplitude is presented in figure 3. The curves in figure 3 illustrate maximal phase error values with respect to constant phase component.

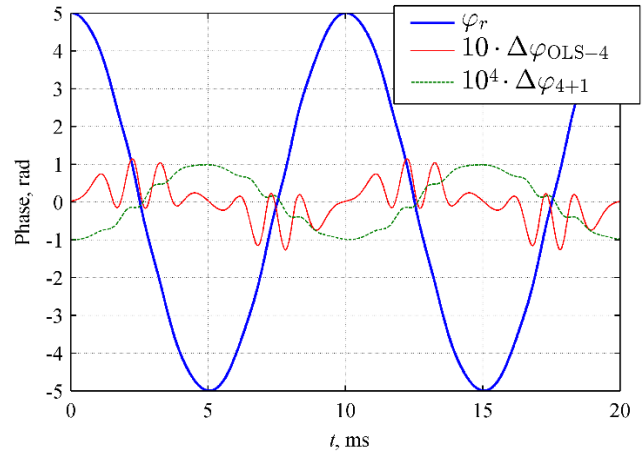

Figure 2. Comparison of target phase signal and errors in case of two algorithms.

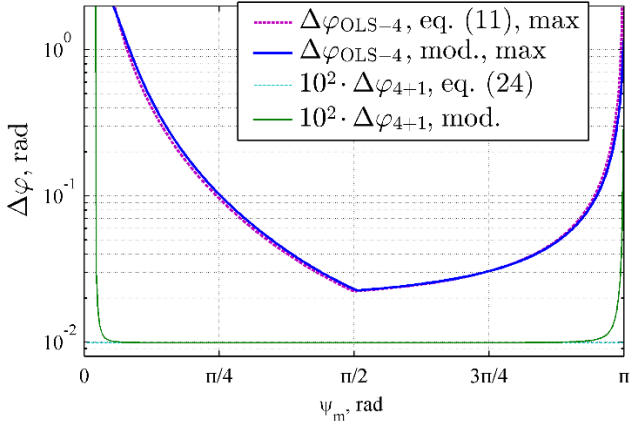

Figure 3. Phase noise magnitude dependency on the amplitude of the phase modulation.

For OLS-4 method, maximal error value with respect to constant phase $\varphi_0$ is presented, while for 4+1 method the error is independent of the $\varphi_0$.

As can be seen in figure 3, there is a good agreement between theoretical estimations according to eqs. (11) and (24) and results of numeric modeling, which proves the correctness of our derivations. However, despite the fact that demodulation error for 4+1 algorithm is constant and very close to theoretical prediction for most of modulation amplitude values, in case of $\psi_m$ close to zero and $\pi$, resultant error substantially increases due to degeneracy of SLE eq. (2).

However, as well as providing two-three order of magnitude lower error, 4+1 algorithm is efficient in case of much wider range of modulation amplitudes ([0.1 3] radians for 4+1 algorithm vs. [1.3 2.4] radians for OLS-4 algorithm).

## 7. Summary

Initially, in this article we assumed to choose method (6) of phase signal demodulation as main one. The principal considered problem arises due to the fact that signal phase in eq. (6) is considered to be constant on demodulation interval. However, target oscillations of phase $\varphi(t)$ disturb phase constancy and increase of frequency or amplitude of these oscillations will lead to sufficient phase variation on demodulation interval. In order correctly take into account this distortion mechanism, an equation (11), describing the resultant phase error in case of phase demodulation based on eq. (6) was derived. Moreover, a generalized approach for synthesis of algorithms with improved immunity to dynamic distortion was utilized to synthesize a modified algorithm given by eq. (21) of 4+1 form (with $N=4$ and $Q=5$). The level of residual error caused by dynamic distortion was also analytically estimated for 4+1 algorithm. The obtained equations for $\Delta\varphi$ allow to evaluate possible distortions in case of given signal or to evaluate applicable parameters of target signals and estimate applicability of algorithms (6) or (21). The described approaches for error analysis and algorithm synthesis can be applied in case of other values of $N$ and $Q$.

However, both equations (6) and (21) were obtained under assumption of phase modulation with given initial phase shift $\theta=0$ and amplitude $\psi_m$ (both methods will work correctly in case of arbitrary, yet known value of $\psi_m$). For real-world devices these conditions are satisfied with finite accuracies and in case of deviation of real values of phase shift from 0 and modulation amplitude from a given value $\psi_m$ (used in (6) or (21)), distortions of demodulated signal can arise. These errors can be considered by analogy with dynamic perturbations, however, phase modulation is formed in modulation hardware and fluctuation of its parameters is technical problem, which can be handled by corresponding means. Moreover, an auxiliary feedback, including analysis of interference signal parameters for $\theta$ and $\psi_m$ estimation and control of modulator's driving signal. Therefore, the influence of these error sources can be suppressed, although theoretical analysis is relevant due to necessity to define the requirements to modulation hardware in case of application of one of the developed demodulation algorithms. Nevertheless, these questions are not considered in the current article since they can be reduced by technical means, while dynamic perturbation is inevitable error source, which is unknown and cannot be controlled by any means.

Moreover, for practical applications of algorithm (21) as alternative to algorithm (6), it is desirable to estimate the noise level of demodulated signal for both of them in case of equal noise levels of initial interference signals under processing. However, this question also falls out of the scope of this article and requires individual analysis.

In such a way, this article is dedicated to the analysis of dynamic distortion of demodulated phase. An equation for the error arising in case of calculation according to algorithm (6) is derived. A modified algorithm (21) with reduced dynamic distortion of demodulated phase is synthesized. Its efficiency in terms of error suppression, as well as the correctness of the expressions, describing the error magnitude for OLS-4 and 4+1 algorithms, obtained analytically is proved by means of numeric simulation.

## Appendix 1

In [24] general solution of SLE in eq. (2) by means of least-squares fitting is considered in detail for case of linear phase modulation. This method can be generalized for the case of arbitrary $\psi^{(q)}$ by means of introducing new variables $U_1 = U_m\cos(\varphi)$ and $U_2 = -U_m\sin(\varphi)$, in terms of which interference signal is expressed as $u^{(q)} = U_0 + U_1\cos[\psi^{(q)}] + U_2\sin[\psi^{(q)}]$. $U_0$, $U_1$ and $U_2$ can be found by minimizing the mean squared error

$$\varepsilon = \sum_{q=0}^{Q-1} \left(U_0 + U_1\cos(\psi^{(q)}) + U_2\sin(\psi^{(q)}) - u^{(q)}\right)^2 \quad \textbf{(A1-1)}$$

Finding derivatives of eq. (A1-1) with respect to unknown parameters and setting them to zero, one obtains the following system of equations

$$\begin{cases} \sum_{q=0}^{Q-1}\left(U_0 + U_1\cos(\psi^{(q)}) + U_2\sin(\psi^{(q)}) - u^{(q)}\right) = 0; \\ \sum_{q=0}^{Q-1}\left(U_0 + U_1\cos(\psi^{(q)}) + U_2\sin(\psi^{(q)}) - u^{(q)}\right)\cdot\cos(\psi^{(q)}) = 0; \\ \sum_{q=0}^{Q-1}\left(U_0 + U_1\cos(\psi^{(q)}) + U_2\sin(\psi^{(q)}) - u^{(q)}\right)\cdot\sin(\psi^{(q)}) = 0. \end{cases}$$

**(A1-2)**

Analysis of this system of equations allows to find the ratio $U_1/U_2$ and therefore obtain the expression for target phase as

$$\varphi_r = -\mathrm{atan2}\left\{\frac{\sum_{q=0}^{Q-1} u^{(q)}\left[A_{11} + A_{12}\cos(\psi^{(q)}) + A_{13}\sin(\psi^{(q)})\right]}{\sum_{q=0}^{Q-1} u^{(q)}\left[A_{21} + A_{22}\cos(\psi^{(q)}) + A_{23}\sin(\psi^{(q)})\right]}\right\},$$

**(A1-3)**

where

$$A_{11} = \sum_{q=0}^{Q-1} \cos(\psi^{(q)}) \cdot \sum_{q=0}^{Q-1} \cos(\psi^{(q)}) \sin(\psi^{(q)}) - \sum_{q=0}^{Q-1} \sin(\psi^{(q)}) \cdot \sum_{q=0}^{Q-1} \cos^2(\psi^{(q)}),$$
$$A_{12} = \sum_{q=0}^{Q-1} \cos(\psi^{(q)}) \cdot \sum_{q=0}^{Q-1} \sin(\psi^{(q)}) - Q \cdot \sum_{q=0}^{Q-1} \cos(\psi^{(q)}) \sin(\psi^{(q)}),$$
$$A_{13} = Q \cdot \sum_{q=0}^{Q-1} \cos^2(\psi^{(q)}) - \left[\sum_{q=0}^{Q-1} \cos(\psi^{(q)})\right]^2$$
$$A_{21} = \sum_{q=0}^{Q-1} \sin(\psi^{(q)}) \cdot \sum_{q=0}^{Q-1} \cos(\psi^{(q)}) \sin(\psi^{(q)}) - \sum_{q=0}^{Q-1} \cos(\psi^{(q)}) \cdot \sum_{q=0}^{Q-1} \sin^2(\psi^{(q)}),$$
$$A_{22} = Q \cdot \sum_{q=0}^{Q-1} \sin^2(\psi^{(q)}) - \left[\sum_{q=0}^{Q-1} \sin(\psi^{(q)})\right]^2,$$
$$A_{23} = \sum_{q=0}^{Q-1} \cos(\psi^{(q)}) \cdot \sum_{q=0}^{Q-1} \sin(\psi^{(q)}) - Q \cdot \sum_{q=0}^{Q-1} \cos(\psi^{(q)}) \sin(\psi^{(q)})$$

**(A1-4)**

Solution based on (A1-3) and (A1-4) is suitable for arbitrary $Q$ and $\psi_m$ under condition of at least three values $\psi^{(q)}$ unequal by modulo $2\pi$.

## Appendix 2

In case of $Q=4$ demodulation period contains four samples with different phases, potentially allowing to develop an algorithm, meeting additional requirements, for example, reducing the phase error component proportional to $\delta$. Then, conditions in eq. (13) can be rewritten in form

$$a_0 + a_1 + a_2 + a_3 = 0, \quad a_0 + a_2 + (a_1 + a_3) \cdot \cos(\psi_m) = 0,$$
$$b_0 + b_1 + b_2 + b_3 = 0, \quad (-b_1 + b_3) \cdot \sin(\psi_m) = 0, \quad \textbf{(A2-1)}$$
$$(-a_1 + a_3) \cdot \sin(\psi_m) = b_0 + b_2 + (b_1 + b_3) \cdot \cos(\psi_m) > 0.$$

Substituting eq. (9) into eq. (12) and making derivations, analogous to the ones made during development of eq. (11), the following expression for phase error $\Delta\varphi$ can be obtained

$$\Delta\varphi = \frac{1}{2} \frac{A \sin(2\varphi_0) + B \cos(2\varphi_0) + D}{b_0 + b_2 + (b_1 + b_3) \cos(\psi_m)}, \quad \textbf{(A2-2)}$$

where

$$A = \delta \cdot \left[\frac{3}{2} a_0 + \frac{1}{2} a_1 \cos(\psi_m) - \frac{1}{2} a_2 - \frac{3}{2} a_3 \cos(\psi_m) - \frac{1}{2} b_1 \sin(\psi_m) - \frac{3}{2} b_3 \sin(\psi_m)\right],$$
$$B = \delta \cdot \left[\frac{1}{2} a_1 \sin(\psi_m) + \frac{3}{2} a_3 \sin(\psi_m) + \frac{3}{2} b_0 + \frac{1}{2} b_1 \cos(\psi_m) - \frac{1}{2} b_2 - \frac{3}{2} b_3 \cos(\psi_m)\right],$$
$$D = \delta \cdot \left[\frac{1}{2} a_1 \sin(\psi_m) + \frac{3}{2} a_3 \sin(\psi_m) - \frac{3}{2} b_0 - \frac{1}{2} b_1 \cos(\psi_m) + \frac{1}{2} b_2 + \frac{3}{2} b_3 \cos(\psi_m)\right].$$

**(A2-3)**

In order to obtain noise-reduction algorithm, values $a_q$ and $b_q$, providing $A=B=D=0$ and fulfilling conditions in eq. (A2-1). In such a manner, we get eight equations and seven variables, however, in this particular case all eight conditions can't be fulfilled simultaneously (for example, $B$ and $D$ can't be simultaneously set to zero).

This means that error compensating algorithm in case of harmonic modulation and $N=Q=4$ is impossible. This reflects the fact that redundancy of samples on demodulation interval alone doesn't enable one to develop a demodulation algorithm with arbitrary additional requirements. As like as not, compensation of error, induced by linear phase increment can be performed in case of modulation signal of a general form in eq. (3) with $\theta \neq 0$, but this case is out of scope of the current work. Moreover, having 4 samples of the same form as eq. (9), enables compensation of another error source – for example, of small parasitic deviation of $\theta$ from zero, however, this question also falls out of scope of this paper.

## Appendix 3

In derivation of eqs. (10) and (11) expressions for coefficients $S$, $C$, $S_r$ and $C_r$ are needed, as well as their derivatives with respect to $\delta$. From eqs. (6) and (9), $S_r$ and $C_r$ can be written as

$$S_r = -\frac{1-\cos(\psi_m)}{\sin(\psi_m)} \cdot \left[\cos\left(\varphi_0 - \frac{1}{2}\delta + \psi_m\right) - \cos\left(\varphi_0 + \frac{3}{2}\delta - \psi_m\right)\right],$$
$$C_r = \cos\left(\varphi_0 - \frac{3\delta}{2}\right) - \cos\left(\varphi_0 - \frac{\delta}{2} + \psi_m\right) + \cos\left(\varphi_0 + \frac{\delta}{2}\right) - \cos\left(\varphi_0 + \frac{3\delta}{2} - \psi_m\right).$$

**(A3-1)**

In case of constant target phase during the demodulation interval, equal to $\varphi_0$, coefficients $S$ and $C$ can be obtained from eq. (A3-1) by substituting $\delta=0$ and making trivial simplifications

$$S = 2[1 - \cos(\psi_m)] \cdot \sin(\varphi_0),$$
$$C = 2[1 - \cos(\psi_m)] \cos(\varphi_0). \quad \textbf{(A3-2)}$$

Derivatives of $S_r$ and $C_r$ with respect to $\delta$ can be written as

$$S' = \left.\frac{dS_r}{d\delta}\right|_{\delta=0} = -\frac{1-\cos(\psi_m)}{\sin(\psi_m)} (2\sin(\varphi_0)\cos(\psi_m) - \cos(\varphi_0)\sin(\psi_m)),$$
$$C' = \left.\frac{dC_r}{d\delta}\right|_{\delta=0} = \sin(\varphi_0) + \sin(\varphi_0)\cos(\psi_m) - 2\cos(\varphi_0)\sin(\psi_m).$$

**(A3-3)**

Coefficients for 4+1 algorithm are different and analysis requires their second derivatives with respect to $\delta$. In this case, taking into account eq. (12), $S_r$ and $C_r$ can be written in the following form

$$S_r = \sum_{q=0}^{4} a_q u^{(q)}, \quad C_r = \sum_{q=0}^{4} b_q u^{(q)}. \quad \textbf{(A3-4)}$$

Substituting eq. (14) into eq. (A3-4) and using eq. (15), one obtains for case of $\delta=0$

$$S = (-a_1 + a_3)\sin(\psi_m)\sin(\varphi_0),$$
$$C = (-a_1 + a_3)\sin(\psi_m)\cos(\varphi_0), \quad \textbf{(A3-5)}$$

from where one can find first derivatives

$$S' = 2(a_0 - a_4)\sin(\varphi_0) + (a_1 - a_3)\sin(\varphi_0)\cos(\psi_m)$$
$$+ (a_1 + a_3)\cos(\varphi_0)\sin(\psi_m),$$
$$C' = 2(b_0 - b_4)\sin(\varphi_0) + (b_1 - b_3)\sin(\varphi_0)\cos(\psi_m) \quad \textbf{(A3-6)}$$
$$+ (b_1 + b_3)\cos(\varphi_0)\sin(\psi_m).$$

Second term in expansion eq. (16) can be written as

$$\left.\frac{d^2}{d\delta^2} \text{atan2}\left(\frac{S_r}{C_r}\right)\right|_{\delta=0} = \frac{(S''C - C''S)(C^2 + S^2) - 2(CC' + SS')(S'C - C'S)}{(C^2 + S^2)^2}$$

**(A3-7)**

where second derivatives with respect to $\delta$ can be expressed as

$$S'' = \left.\frac{d^2 S_r}{d\delta^2}\right|_{\delta=0} = -4(a_0+a_4)\cos(\varphi_0) - (a_1+a_3)\cos(\varphi_0)\cos(\psi_m) + (a_1-a_3)\sin(\varphi_0)\sin(\psi_m),$$

$$C'' = \left.\frac{d^2 C_r}{d\delta^2}\right|_{\delta=0} = -4(b_0+b_4)\cos(\varphi_0) - (b_1+b_3)\cos(\varphi_0)\cos(\psi_m) + (b_1-b_3)\sin(\varphi_0)\sin(\psi_m).$$

(A3-8)

It should be noted that in equations above the factor $U_m$ is omitted since it can be cancelled out being inclusive in both nominator and denominator of atan function argument.

**Funding Information.** The work was done under financial support of Ministry of Education and Science of the Russian Federation in terms of FTP "Research and development on priority trends of Russian scientific-technological complex evolvement in 2014-2020 years (agreement # 14.578.21.0211, agreement unique identifier RFMEFI57816X0211)".

## References

1. J. Cole, C. K. Kirkendall, A. Dandridge, G. Cogdell, and T. G. Giallorenzi, "Twenty-five years of interferometric fiber optic acoustic sensors at the naval research laboratory," J. Washingt. Acad. Sci. 90, 40–56 (2004).
2. K. Creath, "Phase-measurement interferometry techniques," Prog. Opt. 24, 349–393 (1988).
3. H. Schreiber and J. H. Bruning, "Phase shifting interferometry," in Optical Shop Testing (2007), p. 547 — 666.
4. O. I. Kotov, L. B. Liokumovich, S. I. Markov, A. V. Medvedev, and V. M. Nikolaev, "Remote interferometer with polarizing beam splitting," Tech. Phys. Lett. 26, 415–417 (2000).
5. L. B. Liokumovich, A. V. Medvedev, and V. Petrov, "Fiber-optic polarization interferometer with an additional phase modulation for electric field measurements," Opt. Mem. Neural Networks (Information Opt. 22, 21–27 (2013).
6. H.P. Stahl, "Review of phase-measuring interferometry", in Optical Testing and Metrology III: Recent Advances in Industrial Optical Detection, C.P. Grover, ed., Proc. Soc. Opt. Instrum. Eng. 1332, 704-709 (1991).
7. J. Schwider, R. Burov, K-E Elssner, J. Grzanna, R. Spolackzyc, and K. Merkel, "Digital wave-front measuring interferometry: some systematic error sources, App. Opt. 22, 3421-3432 (1983).
8. D. A. Jackson, A. D. Kersey, A. Corke, and J. D. C. Jones, "Pseudoheterodyne detection scheme for optical interferometers," Electron. Lett. 18, 1081–1083 (1982).
9. A. B. Lobo Ribeiro, R. F. Caleya, and J. L. Santos, "General error function of synthetic-heterodyne signal processing in interferometric fibre-optic sensors," Int. J. Optoelectron. 10, 1081–1083 (1995).
10. A. Dandridge, A. B. Tveten, and T. G. Giallorenzi, "Homodyne demodulation scheme for fiber optic sensors using phase generation generated carrier," IEEE Trans. Microw. Theory Tech. 30, 1635–1641 (1982).
11. C.P. Brophy, "Effect of intensity error correlation on the computed phase of phase-shifting interferometry", J. Opt. Soc. Am. 7, 537-541 (1990).
12. W. Phillion, "General methods for generating phase-shifting interferometry algorithms", App. Opt. 36, 8098-8115 (1997).
13. Q. Wang, "Fourier analysis of phase-shifting algorithms for amplitude measurement of interference fringe," Appl. Opt. 56, 4353 (2017).
14. G. A. Ayubi, C. D. Perciante, J. M. Di Martino, J. L. Flores, and J. a. Ferrari, "Generalized phase-shifting algorithms: error analysis and minimization of noise propagation," Appl. Opt. 55, 1461 (2016).
15. E. Hack and J. Burke, "Measurement uncertainty of linear phase-stepping algorithms," Rev. Sci. Instrum. 82, 61101 (2011).
16. D. Malacara-Hernandez and D. Malacara-Doblado, "Optical Testing and Interferometry," Prog. Opt. **62**, 73–156 (2017).
17. Z. Chen, Y. Liu, H. Li, and M. Yu, "Real-time demodulation scheme based on phase-shifting interferometry with error compensations for miniature Fabry-Perot acoustic sensors," in Proceedings of SPIE (2006), pp. 61670N.
18. Y. Ge, "A Modulation Depth Calibration in Orthogonal Demodulation for Optical Fiber Interferometric Sensor," in The Ninth International Conference on Electronic Measurement & Instruments ICEMI (2009), pp. 2-854-2–858.
19. O. Sasaki, H. Okazaki, and M. Sakai, "Sinusoidal phase modulating interferometer using the integrating-bucket method", App. Opt. 26, 1089-1093 (1987).
20. A. Dubois, "Phase-map measurements by interferometry with sinusoidal phase modulation and four integrating buckets," J. Opt. Soc. Am. A. Opt. Image Sci. Vis. 18, 1972–1979 (2001).
21. P. Hariharan, B.F. Oreb, and T. Eiju, "Digital phase-shifting interferometry: a simple error compensating phase calculating algorithm", App. Opt. 26, 2504-2506 (1987).
22. H. Bi, Y. Zhang, K. V. Ling, and C. Wen, "Class of 4 + 1-phase algorithms with error compensation," Appl. Opt. 43, 4199–4207 (2004).
23. Y. Surrel, "Phase stepping: a new self-calibrating algorithm," Appl. Opt. 32, 3598–600 (1993).
24. D. Malacara, M. Servin, and Z. Malacara, Interferogram Analysis for Optical Testing, Second Edition (Taylor&Francis, 2005).